# Trouble with the Lorentz law of force: Incompatibility with special relativity and momentum conservation

Masud Mansuripur

College of Optical Sciences, The University of Arizona, Tucson, Arizona 85721

masud@optics.arizona.edu

**Abstract**. The Lorentz law of force is the fifth pillar of classical electrodynamics, the other four being Maxwell's macroscopic equations. The Lorentz law is the universal expression of the force exerted by electromagnetic fields on a volume containing a distribution of electrical charges and currents. If electric and magnetic dipoles also happen to be present in a material medium, they are traditionally treated by expressing the corresponding polarization and magnetization distributions in terms of bound-charge and bound-current densities, which are subsequently added to free-charge and free-current densities, respectively. In this way, Maxwell's macroscopic equations are reduced to his microscopic equations, and the Lorentz law is expected to provide a precise expression of the electromagnetic force density on material bodies at all points in space and time. This paper presents incontrovertible theoretical evidence of the incompatibility of the Lorentz law with the fundamental tenets of special relativity. We argue that the Lorentz law must be abandoned in favor of a more general expression of the electromagnetic force density, such as the one discovered by A. Einstein and J. Laub in 1908. Not only is the Einstein-Laub formula consistent with special relativity, it also solves the long-standing problem of "hidden momentum" in classical electrodynamics.

**1. Introduction**. The classical theory of electrodynamics is based on Maxwell's macroscopic equations [1-3], which, in the MKSA system of units, may be written as follows:

$$\nabla \cdot \boldsymbol{D}(\boldsymbol{r},t) = \rho_{\text{free}}(\boldsymbol{r},t), \tag{1}$$

$$\nabla \times \boldsymbol{H}(\boldsymbol{r},t) = \boldsymbol{J}_{\text{free}}(\boldsymbol{r},t) + \partial \boldsymbol{D}(\boldsymbol{r},t)/\partial t, \tag{2}$$

$$\nabla \times \boldsymbol{E}(\boldsymbol{r},t) = -\partial \boldsymbol{B}(\boldsymbol{r},t)/\partial t, \tag{3}$$

$$\nabla \cdot \boldsymbol{B}(\boldsymbol{r},t) = 0. \tag{4}$$

Here $(\boldsymbol{r},t)$ represents position in space-time, $\rho_{\text{free}}$ and $\boldsymbol{J}_{\text{free}}$ are the densities of free charge and free current, $\boldsymbol{E}$ is the electric field, $\boldsymbol{H}$ the magnetic field, $\boldsymbol{D}$ the displacement, and $\boldsymbol{B}$ the magnetic induction. By definition, $\boldsymbol{D}(\boldsymbol{r},t) = \varepsilon_\text{o}\boldsymbol{E}(\boldsymbol{r},t) + \boldsymbol{P}(\boldsymbol{r},t)$, where $\varepsilon_\text{o}$ is the permittivity of free space and $\boldsymbol{P}$ the electric polarization. Similarly, $\boldsymbol{B}(\boldsymbol{r},t) = \mu_\text{o}\boldsymbol{H}(\boldsymbol{r},t) + \boldsymbol{M}(\boldsymbol{r},t)$, where $\mu_\text{o}$ is the permeability of free space and $\boldsymbol{M}$ the magnetization. The speed of light in vacuum is found from these equations to be related to the electromagnetic (EM) properties of free space via $c = 1/\sqrt{\mu_\text{o}\varepsilon_\text{o}}$.

In general, $\rho_{\text{free}}$ and $\boldsymbol{J}_{\text{free}}$ are arbitrary functions of space-time that satisfy the continuity equation $\nabla \cdot \boldsymbol{J}_{\text{free}} + \partial\rho_{\text{free}}/\partial t = 0$. Moreover, the 4-vector $(\boldsymbol{J}_{\text{free}}, c\rho_{\text{free}})$ is readily Lorentz transformed between different inertial frames. The remaining sources of the EM field, $\boldsymbol{P}$ and $\boldsymbol{M}$, are also arbitrary functions of space-time which collectively form a 2nd rank tensor that can be Lorentz transformed from one inertial frame to another. Two other 2nd rank tensors that obey the Lorentz transformation rules are the field tensors, one formed by $\boldsymbol{E}$ and $\boldsymbol{B}$, the other by $\boldsymbol{D}$ and $\boldsymbol{H}$ [1,3,4].

One could define bound-charge and bound-current densities $\rho_{\text{bound}}(r,t) = -\nabla \cdot P$ and $J_{\text{bound}}(r,t) = (\partial P/\partial t) + \mu_o^{-1} \nabla \times M$, and use them to eliminate $D$ and $H$ from Eqs.(1-4). Maxwell's equations then reduce to the so-called microscopic equations, relating $E$ and $B$ fields to the total charge- and current-densities $\rho_{\text{total}} = \rho_{\text{free}} + \rho_{\text{bound}}$ and $J_{\text{total}} = J_{\text{free}} + J_{\text{bound}}$. It is thus clear that, as far as Maxwell's equations are concerned, $P(r,t)$ and $M(r,t)$ can be treated as distributions of charge and current, and that the knowledge of $\rho_{\text{total}}(r,t)$ and $J_{\text{total}}(r,t)$ is all that is needed to determine the corresponding EM fields throughout the entire space-time.

$P(r,t)$ and $M(r,t)$ cease to behave as mere distributions of charge and current, however, as soon as one takes a close look at their interactions with EM fields involving energy and momentum exchange [5]. Since we have discussed these issues at great length elsewhere [6-12], we confine our remarks here to a summary of the conclusions reached in earlier studies, namely,

i) If the magnetic dipoles of Maxwell's equations were ordinary current loops, the rate of flow of EM energy (per unit area per unit time), instead of being the commonly accepted Poynting vector $S(r,t) = E \times H$ [2,3,5], would be given by $\mu_o^{-1} E \times B$ [1].

ii) If the force-density exerted by EM fields on material media obeyed the conventional Lorentz law, namely,

$$F(r,t) = \rho_{\text{total}} E + J_{\text{total}} \times B, \qquad (5)$$

there would arise situations where the momentum of a closed system would not be conserved. This problem has been known since the 1960s, when W. Shockley [13,14] pointed out the existence of "hidden momentum" in certain electromagnetic systems [15,16].

iii) A generalized version of the Lorentz law, originally proposed in 1908 by A. Einstein and J. Laub [17,18] and independently rediscovered by several authors afterward [5,16,19-21], not only justifies the definition of the Poynting vector as $S(r,t) = E \times H$, but also eliminates the problem of hidden momentum, thus bringing classical electrodynamics into compliance with momentum conservation laws [10,15,20]. The Einstein-Laub formula for force density is

$$F(r,t) = \rho_{\text{free}} E + J_{\text{free}} \times \mu_o H + (P \cdot \nabla) E + (\partial P/\partial t) \times \mu_o H + (M \cdot \nabla) H - (\partial M/\partial t) \times \varepsilon_o E. \qquad (6)$$

To guarantee the conservation of angular momentum, Eq.(6) must be supplemented with the following expression for the torque-density exerted by EM fields on material media:

$$T(r,t) = r \times F(r,t) + P(r,t) \times E(r,t) + M(r,t) \times H(r,t). \qquad (7)$$

Equations (6) and (7) guarantee momentum conservation under all circumstances provided that the Abraham momentum-density, $p(r,t) = S(r,t)/c^2$, is assumed for EM momentum, and that the EM angular-momentum-density is accordingly assumed to be $L(r,t) = r \times S(r,t)/c^2$.

iv) Conservation of energy is guaranteed by the Poynting theorem derived from Maxwell's Eqs.(2) and (3), in conjunction with the postulate that the rate and direction of EM energy flow are given by the Poynting vector $S(r,t) = E \times H$. The energy continuity equation may thus be written as

$$\nabla \cdot S(r,t) + \frac{\partial}{\partial t} (\tfrac{1}{2} \varepsilon_o E \cdot E + \tfrac{1}{2} \mu_o H \cdot H) + E \cdot J_{\text{free}} + E \cdot \frac{\partial P}{\partial t} + H \cdot \frac{\partial M}{\partial t} = 0. \qquad (8)$$

Note in the above equation that energy exchange between EM fields and electric dipoles is governed by the term $\boldsymbol{E}\cdot\partial \boldsymbol{P}/\partial t$, which is sensible considering the force term $(\boldsymbol{P}\cdot\boldsymbol{\nabla})\boldsymbol{E}$ of Eq.(6) and the torque term $\boldsymbol{P}\times\boldsymbol{E}$ of Eq.(7). Similarly, the exchange of energy with magnetic dipoles is governed by the term $\boldsymbol{H}\cdot\partial \boldsymbol{M}/\partial t$ of Eq.(8), which is compatible with the force term $(\boldsymbol{M}\cdot\boldsymbol{\nabla})\boldsymbol{H}$ and the torque term $\boldsymbol{M}\times\boldsymbol{H}$ if one imagines a magnetic dipole as a pair of north and south magnetic poles attached to the opposite ends of a short spring.

The goal of the present paper is to demonstrate that the conventional Lorentz law of Eq.(5) violates a basic requirement of special relativity. The elementary example presented in the next section establishes the inadequacy of the Lorentz law; several other examples could be cited in support of the argument, but a single instance of incongruity is all it takes to prove the point. In addition, we show that the Einstein-Laub force-density expression of Eq.(6), taken together with the torque-density expression of Eq.(7), does *not* suffer from the aforesaid shortcoming. The totality of evidence against the Lorentz law and in support of the Einstein-Laub formula thus compels us to abandon the former in favor of the latter. The ultimate test of any physical theory, of course, is whether or not its predictions agree with experimental observations. Thus, our argument in favor of the Einstein-Laub formulation should not be construed as an argument against any other law of force that has been proposed in the past or might be proposed in the future, provided that the proposed law is also universal, is consistent with Maxwell's equations and with conservation laws, and complies with special relativity.

**2. Lorentz law and the principle of relativity**. Consider a point-charge $q$ at a fixed distance $d$ from a magnetic point-dipole $m_\text{o}\hat{\boldsymbol{x}}'$, as shown in Fig.1. The magnetic dipole which, in the Lorentz picture, is essentially a small, charge-neutral loop of current, experiences neither a force nor a torque from the point-charge. The situation is very different, however, for a stationary observer in the $xyz$ frame, who watches the point-charge and the magnetic dipole move together with constant velocity $V$ along the $z$-axis. It is not difficult to Lorentz transform the current loop and also the electric field of the point-charge from $x'y'z'$ to the $xyz$ frame. The stationary observer will see an electric dipole $p_\text{o}\hat{\boldsymbol{y}} = V\varepsilon_\text{o} m_\text{o}\hat{\boldsymbol{y}}$, traveling along with the magnetic dipole $m_\text{o}\hat{\boldsymbol{x}}$, both experiencing the electric as well as the magnetic field produced by the moving point-charge $q$. The net Lorentz force on the pair of dipoles ($m_\text{o}\hat{\boldsymbol{x}}$ and $p_\text{o}\hat{\boldsymbol{y}}$) turns out to be zero, but a net torque $\boldsymbol{T} = (Vqm_\text{o}/4\pi d^2)\hat{\boldsymbol{x}}$ acts on the dipole pair. The appearance of this torque in the $xyz$ frame in the absence of a corresponding torque in the $x'y'z'$ frame is sufficient proof of the inadequacy of the Lorentz law. In contrast, the force- and torque-density expressions in Eqs.(6) and (7) yield zero force and zero torque on the dipole(s) in both reference frames.

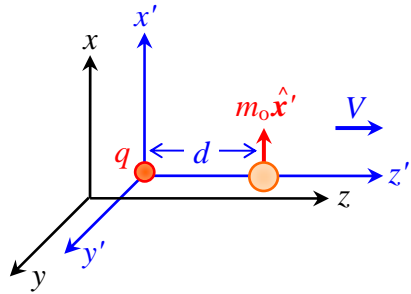

**Fig.1**. In the inertial $x'y'z'$ frame, the point-charge $q$ and the point-dipole $m_\text{o}\hat{\boldsymbol{x}}'$ are stationary. The $x'y'z'$ system moves with constant velocity $V$ along the $z$-axis, as seen by a stationary observer in the $xyz$ frame. The origins of the two coordinate systems coincide at $t = t' = 0$.

The proof of the above statements requires the use of Dirac's delta-function $\delta(\cdot)$ to represent the magnetization of the point-dipole in the $x'y'z'$ frame as

$$\boldsymbol{M}'(\boldsymbol{r}',t') = m_o \hat{\boldsymbol{x}}' \delta(x')\delta(y')\delta(z'-d). \tag{9}$$

Lorentz transformation to the $xyz$ frame then yields a pair of point-dipoles, as follows:

$$\boldsymbol{M}(\boldsymbol{r},t) = \gamma m_o \hat{\boldsymbol{x}} \delta(x)\delta(y)\delta[\gamma(z-Vt)-d], \tag{10a}$$

$$\boldsymbol{P}(\boldsymbol{r},t) = \gamma V \varepsilon_o m_o \hat{\boldsymbol{y}} \delta(x)\delta(y)\delta[\gamma(z-Vt)-d]. \tag{10b}$$

Here $\gamma = 1/\sqrt{1-V^2/c^2}$. In the $x'y'z'$ frame, the $E$-field produced by the point-charge $q$ is

$$\boldsymbol{E}'(\boldsymbol{r}',t') = \frac{q(x'\hat{\boldsymbol{x}}' + y'\hat{\boldsymbol{y}}' + z'\hat{\boldsymbol{z}}')}{4\pi\varepsilon_o(x'^2 + y'^2 + z'^2)^{3/2}}. \tag{11}$$

When the above $E$-field is Lorentz transformed to the $xyz$ frame, the resulting fields will be

$$\boldsymbol{E}(\boldsymbol{r},t) = \frac{\gamma q[x\hat{\boldsymbol{x}} + y\hat{\boldsymbol{y}} + (z-Vt)\hat{\boldsymbol{z}}]}{4\pi\varepsilon_o[x^2 + y^2 + \gamma^2(z-Vt)^2]^{3/2}}, \tag{12a}$$

$$\boldsymbol{H}(\boldsymbol{r},t) = \frac{\gamma V q(x\hat{\boldsymbol{y}} - y\hat{\boldsymbol{x}})}{4\pi[x^2 + y^2 + \gamma^2(z-Vt)^2]^{3/2}}. \tag{12b}$$

To calculate the conventional Lorentz force on the dipole pair of Eqs. (10), we add the force of the $E$-field on bound charges to that of the $B$-field on bound currents; see Eq. (5). The total force and torque will then be found by integrating the corresponding densities, namely,

$$\boldsymbol{F}(\boldsymbol{r},t) = -(\nabla\cdot\boldsymbol{P})\boldsymbol{E} + [(\partial\boldsymbol{P}/\partial t) + \mu_o^{-1}\nabla\times\boldsymbol{M}]\times\mu_o\boldsymbol{H}, \tag{13a}$$

$$\boldsymbol{T}(\boldsymbol{r},t) = \boldsymbol{r}\times\boldsymbol{F}(\boldsymbol{r},t). \tag{13b}$$

The calculation is straightforward with the aid of the sifting property of the delta-function and its derivative, namely,

$$\int_{-\infty}^{\infty} f(x)\delta(x-x_o)\,dx = f(x_o), \tag{14}$$

$$\int_{-\infty}^{\infty} f(x)\delta'(x-x_o)\,dx = -f'(x_o). \tag{15}$$

In similar fashion, one can determine the Einstein-Laub force and the corresponding torque on the dipole pair of Eqs. (10) using the densities given in Eqs. (6) and (7).

We close this section by noting that certain unusual situations could arise when "force" is transformed from one inertial frame to another. For example, in the coordinate system of Fig. 1, a thin, straight, charge-neutral wire carrying a constant, uniform current-density $\boldsymbol{J}_{\text{free}}$ along $x'$ in the presence of a constant, uniform $E$-field (also along $x'$) does *not* experience a Lorentz force. However, the same wire seen by the stationary observer in the $xyz$ frame *does* experience a Lorentz force along the $z$-axis. Nevertheless, special relativity is not violated here, because the energy delivered by the $E$-field at the rate of $\boldsymbol{E}\cdot\boldsymbol{J}_{\text{free}}$ to the current causes an increase in the mass of the wire. Seen by the observer in the $xyz$ frame, the wire has a non-zero (albeit constant) velocity along $z$ and, therefore, its relativistic momentum $\boldsymbol{p}$ increases with time, not because of a change of velocity but because of a change of mass! The observed electromagnetic force in the moving frame thus agrees with the relativistic version of Newton's law, $F = \mathrm{d}p/\mathrm{d}t$. In contrast, the situation depicted in Fig. 1, where a magnetic dipole experiences a torque in the moving frame, does *not* have a similar explanation.

**3. Concluding remarks**. We have presented a single example of a system that violates special relativity under the Lorentz law, while complying with relativity under the Einstein-Laub formula. There exist several other examples that show the superiority of the Einstein-Laub force expression (accompanied by the corresponding torque formula). Considering that the Lorentz law also fails to conserve momentum in certain situations involving magnetic media, while the Einstein-Laub force (and its associated torque) show across-the-board consistency with the conservation laws, it is only natural to consider abandoning the former and adopting the latter as the universal law of electromagnetic force (and torque).

We are cognizant of the fact that the mathematical consistency of the Einstein-Laub force equation with conservation laws and with special relativity does not, in itself, establish the validity of the formula as a law of Nature. Nor does our demonstration of consistency exclude other formulations from being viable candidates to replace (or to augment) the Lorentz law. Ultimately, of course, it is agreement with a broad range of experimental observations that is needed to establish the validity of a given formulation.

The extensive review by Brevik [22] contains a wealth of information about early efforts to determine an appropriate energy-momentum tensor for the electromagnetic field in the presence of dielectric and magnetic media. In some of the reported experiments, the predictions of the Helmholtz force equation were found to be closer to actual observations than those of the Einstein-Laub formulation. Without discounting the possibility that the experimental results were in fact correctly interpreted, we point out that theoretical analyses based on the stress-energy tensor are fraught with errors and inaccuracies. Loudon [23], for example, has emphasized "*the simplicity and safety of calculations based on the Lorentz force and the dangers of calculations based on derived expressions involving elements of the Maxwell stress tensor, whose contributions may vanish in some situations but not in others.*" It is also possible, starting with the Einstein-Laub force and the Abraham momentum density expressions, to arrive at a different stress tensor than the one that has been used in the past to evaluate the predictions of Einstein and Laub against experimental observations [8]. All in all, it is our contention that the existing evidence against the Einstein-Laub formula, while serious, is not fatal, and that its aforementioned strengths along with its widespread resurgence in the literature [5,15,16,19-21] – especially since Shockley's discovery of hidden momentum [13,14] – justify a closer examination of the experimental evidence against the formulation.

Assuming that experimental observations end up supporting the Einstein-Laub hypothesis, the fundamental postulates of classical electrodynamics will be the macroscopic equations of Maxwell, Eqs. (1) to (4), the Einstein-Laub force-density expression, Eq. (6), the torque-density expression, Eq. (7), the Poynting postulate that $S(r,t) = E \times H$ represents the rate of flow of EM energy, and the Abraham postulate that EM momentum and angular momentum densities are given by $p(r,t) = S(r,t)/c^2$ and $L(r,t) = r \times S(r,t)/c^2$, respectively. These postulates, which are consistent with the conservation laws of energy, momentum, and angular momentum, and also comply with the requirements of special relativity, would then apply to any electromagnetic system in which free charge, free current, polarization, and magnetization interact with EM fields.

Note that we have not specified any constitutive relations that would determine the response of material media to EM fields. The fundamental postulates of electrodynamics are completely general and apply to all kinds of material media, be they linear or nonlinear, passive or active, moving or stationary, and whether or not these media are dispersive, absorptive, homogeneous, hysteretic, etc.

We emphasize, once again, that although *P* and *M* can be replaced with equivalent bound-charge and bound-current densities within Maxwell's equations, such substitutions are *not* allowed in the context of force, torque, and energy. In other words, electric and magnetic dipoles exhibit certain special properties in their interactions with EM fields that set them apart from ordinary charge and current distributions. An electric dipole should not be considered as merely a pair of positive and negative charges attached to the opposite ends of a spring, nor should a magnetic dipole be thought of as a simple Amperian current loop. With the demise of the Lorentz law of force, electric and magnetic dipoles acquire individual identities above and beyond what Maxwell's equations have traditionally ascribed to them. The essential quantum mechanical nature of electric and magnetic dipoles is such that their interactions with EM fields cannot be described in terms of equivalent (bound) charge and current densities; rather, these interactions are governed by Eqs.(6) and (7) when linear and angular momenta are being exchanged, and by Eq.(8) in situations involving an exchange of energy.

**Acknowledgements**. The author is grateful to Tobias S. Mansuripur, Ewan M. Wright, Rodney Loudon, and Nader Engheta for illuminating discussions.